\documentclass[journal]{IEEEtran}
\usepackage[T1]{fontenc}
\usepackage{cite}
\usepackage{amsmath,amssymb,amsfonts}
\usepackage{algorithm}
\usepackage{algorithmic}
\usepackage{graphicx}
\usepackage{textcomp}
\usepackage{xcolor}
\usepackage{subfigure}
\usepackage{booktabs}
\usepackage{multirow}
\usepackage{diagbox}
\usepackage{url}
\usepackage{threeparttable}
\usepackage{array}
\usepackage{textcase}
\usepackage{rotating}
\usepackage{etoolbox}

\makeatletter
\def\UrlAlphabet{%
      \do\a\do\b\do\c\do\d\do\e\do\f\do\g\do\h\do\i\do\j%
      \do\k\do\l\do\m\do\n\do\o\do\p\do\q\do\r\do\s\do\t%
      \do\u\do\v\do\w\do\x\do\y\do\z\do\A\do\B\do\C\do\D%
      \do\E\do\F\do\G\do\H\do\I\do\J\do\K\do\L\do\M\do\N%
      \do\O\do\P\do\Q\do\R\do\S\do\T\do\U\do\V\do\W\do\X%
      \do\Y\do\Z}
\def\UrlDigits{\do\1\do\2\do\3\do\4\do\5\do\6\do\7\do\8\do\9\do\0}
\g@addto@macro{\UrlBreaks}{\UrlOrds}
\g@addto@macro{\UrlBreaks}{\UrlAlphabet}
\g@addto@macro{\UrlBreaks}{\UrlDigits}

\patchcmd{\@algocf@start}
  {-1.5em}
  {0pt}
  {}{}

\makeatother

\newcommand{\Times}[2]{${\text{#1}\times\text{#2}}$}

\graphicspath{{figures/}}

\begin{document}
\setlength{\textfloatsep}{5pt}
\setlength{\floatsep}{5pt}
\ifdefined \GramaCheck
  \newcommand{\CheckRmv}[1]{}
  \newcommand{\figref}[1]{Figure 1}%
  \newcommand{\tabref}[1]{Table 1}%
  \newcommand{\secref}[1]{Section 1}
  \newcommand{\algref}[1]{Algorithm 1}
  \renewcommand{\eqref}[1]{Equation 1}
\else
  \newcommand{\CheckRmv}[1]{#1}
  \newcommand{\figref}[1]{Fig.~\ref{#1}}%
  \newcommand{\tabref}[1]{Table~\ref{#1}}%
  \newcommand{\secref}[1]{Sec.~\ref{#1}}
  \newcommand{\algref}[1]{Algorithm~\ref{#1}}
  \renewcommand{\eqref}[1]{(\ref{#1})}
\fi
\newtheorem{theorem}{Theorem}
\newtheorem{proposition}{Proposition}
\newtheorem{assumption}{Assumption}
\newtheorem{definition}{Definition}
\newtheorem{condition}{Condition}
\newtheorem{property}{Property}
\newtheorem{remark}{Remark}
\newtheorem{lemma}{Lemma}
\newtheorem{corollary}{Corollary}
%

\title{Next-Generation AI-Native Wireless Communications: MCMC-Based Receiver Architectures for Unified Processing}

%
%
\author{Xingyu~Zhou,~\IEEEmembership{Graduate Student Member,~IEEE,}
        Le~Liang,~\IEEEmembership{Member,~IEEE,}
        Jing~Zhang,~\IEEEmembership{Member,~IEEE,}
		    Chao-Kai~Wen,~\IEEEmembership{Fellow,~IEEE,}
        and~Shi~Jin,~\IEEEmembership{Fellow,~IEEE}
\thanks{X.~Zhou, L. Liang, J.~Zhang, and S.~Jin are with the National Mobile
Communications Research Laboratory, Southeast University, Nanjing 210096, China
(e-mail: \protect \url{xy_zhou@seu.edu.cn}; lliang@seu.edu.cn; jingzhang@seu.edu.cn; jinshi@seu.edu.cn). L. Liang is also with Purple Mountain Laboratories, Nanjing 211111, China.}
\thanks{C.-K.~Wen is with the Institute of Communications Engineering,
National Sun Yat-sen University, Kaohsiung 80424, Taiwan
(e-mail: chaokai.wen@mail.nsysu.edu.tw).}
}

%
%

\maketitle

\begin{abstract}
The multiple-input multiple-output (MIMO) receiver processing is a key technology for current and next-generation wireless communications. However, it faces significant challenges related to complexity and scalability as the number of antennas increases. Artificial intelligence (AI), a cornerstone of next-generation wireless networks, offers considerable potential for addressing these challenges. This paper proposes an AI-driven, universal MIMO receiver architecture based on Markov chain Monte Carlo (MCMC) techniques. Unlike existing AI-based methods that treat receiver processing as a black box, our MCMC-based approach functions as a generic Bayesian computing engine applicable to various processing tasks, including channel estimation, symbol detection, and channel decoding. This method enhances the interpretability, scalability, and flexibility of receivers in diverse scenarios. Furthermore, the proposed approach integrates these tasks into a unified probabilistic framework, thereby enabling overall performance optimization. This unified framework can also be seamlessly combined with data-driven learning methods to facilitate the development of fully intelligent communication receivers.
\end{abstract}


%
\IEEEpeerreviewmaketitle

\vspace{-0.5cm}
\section{Introduction}  
\IEEEPARstart{T}{he} upcoming International Mobile Telecommunications-2030 (IMT-2030), commonly known as 6G, envisions ultra-high data rates, low latency, and enhanced quality of service. These diverse requirements, combined with increased network complexity and heterogeneity, impose significant challenges on system design. Recently, artificial intelligence (AI) has demonstrated notable potential in addressing issues previously considered difficult due to prohibitive computational complexity or the absence of accurate models, prompting transformative shifts in wireless system design principles.

Unlike previous generations of communication networks, in which AI played a supporting role, 6G aspires to build an AI-native wireless network. This ambition requires a framework that seamlessly integrates communication, computation, and AI-driven technologies, ultimately enabling unprecedented levels of performance, autonomy, and intelligence. One key area poised to benefit from this integration is multiple-input multiple-output (MIMO) receiver processing—a fundamental component of modern wireless communications.
 
Wireless communication systems are increasingly moving toward the deployment of ultra-massive antenna arrays to support extremely high data throughput and extended coverage. This trend leads to high-dimensional signal-processing tasks at the receiver, which in turn requires substantial increases in computing capability and power consumption \cite{wang2024tutorial}. Traditional MIMO receiver architectures struggle to meet the escalating complexity and scalability requirements of future communication networks. In this context, AI-based methodologies are emerging as promising solutions \cite{qin2024ai}.

Recent advances in AI-based receivers have focused on using data-driven techniques to enhance MIMO performance. However, these methods often lack transparency, interpretability, and generalizability, particularly with regard to the probabilistic nature of wireless propagation. In fluctuating channel environments, data-driven receivers tend to lack robustness and require frequent retraining, which becomes problematic in real-time scenarios where rapid adaptability is essential. Furthermore, such receivers typically rely on overparameterized models that demand substantial labeled training datasets, resulting in computational and storage overhead that challenges the hardware constraints of standard communication equipment. These limitations underscore the need for a more principled and scalable approach to MIMO receiver design.
  
Markov chain Monte Carlo (MCMC) methods \cite{brooks2011handbook} offer distinct advantages in this context. As a powerful probabilistic sampling technique, MCMC naturally aligns with the Bayesian inference framework that underpins many signal-processing tasks in wireless communications. Recent advancements in MCMC, particularly the integration with machine learning techniques such as gradient-based approaches \cite{maSamplingCanBe2019} and deep generative networks, have significantly enhanced its efficiency for high-dimensional inference problems. Moreover, unlike conventional AI-based models, MCMC methods provide high interpretability through the Bayesian framework, rendering them well-suited for dynamic scenarios and fluctuating channel conditions where robust generalization is critical. 

In addition, MCMC offers significant advantages in hardware implementation due to its ability to exploit massive parallel computing. By distributing sampling tasks across independent, lightweight computing units, MCMC can efficiently scale without relying on high-end centralized processing hardware. This modularity, combined with structured computational design and advanced machine learning techniques, holds great promise for establishing an AI-native processing architecture for large-scale MIMO receivers.
 
In this paper, we introduce MCMC as an efficient probabilistic solver for the challenges encountered in MIMO receiver processing, including channel estimation, symbol detection, and channel decoding. Beyond addressing individual tasks, we investigate the potential of MCMC to unify the receiver pipeline into a comprehensive framework that integrates seamlessly with data-driven approaches. Finally, we discuss future research directions for leveraging MCMC in next-generation communication networks.

 
\vspace{-0.3cm}
\section{Basics of MCMC}
\subsection{System Model and Receiver Processing}

Equipping communication systems with multiple transmit and receive antennas introduces additional degrees of freedom for spatially multiplexing multiple data streams over the MIMO channel, thereby significantly enhancing system capacity. In a typical MIMO transmission, the message bits $\mathbf{b}$ are encoded and modulated to obtain the transmitted symbols. The symbols from multiple transmit antennas collectively form the transmitted vector $\mathbf{x}$. This vector undergoes multipath fading and is corrupted by additive noise during transmission, resulting in the received signal $\mathbf{y} = \mathbf{Hx} + \mathbf{n}$, 
where $\mathbf{H}$ is the channel matrix that represents the path gains between each transmit-receive antenna pair, and $\mathbf{n}$ denotes the noise vector.

The receiver processing chain typically consists of three core tasks: channel estimation, symbol detection, and channel decoding, all of which can be viewed as posterior inference problems from a Bayesian perspective. The goal is to estimate the unknowns ($\mathbf{H}$, $\mathbf{x}$, or $\mathbf{b}$) based on the observation $\mathbf{y}$ and additional side information. This process becomes particularly challenging as communication links grow complicated, resulting in high dimensionality and nonlinearity in the receiver processing tasks.

\vspace{-0.4cm}
\subsection{The Need for MCMC in Receiver Processing}  
Traditional optimal receiver processing relies on criteria such as maximum a posteriori (MAP) or maximum likelihood (ML) estimation. However, these statistical estimates often lack closed-form solutions and require solving high-dimensional integrals or optimization problems, which becomes computationally intractable in large-scale MIMO systems and complex channel environments.

To address these challenges, deterministic approximations are frequently used, but they are often criticized for their lack of precision and robustness. In contrast, stochastic methods, such as Monte Carlo-based sampling, provide an alternative approach by generating a large number of samples from the distributions of interest. This enables more flexible and accurate inference, particularly in high-dimensional spaces.

Compared to traditional receivers and existing AI-aided receivers, MCMC-based receivers provide key advantages in scalability and interpretability. Their integration of domain knowledge enhances the generalization capabilities of the receiver across diverse communication environments. Moreover, MCMC can be employed as a fundamental signal processing block, specifically designed and optimized to efficiently manage various processing tasks. This advantage facilitates the integration of different receiver processing tasks into a unified probabilistic framework, seamlessly combining model-based optimization with neural network (NN) enhancements to create a comprehensive AI-driven processing pipeline.

\vspace{-0.4cm}
\subsection{Overview of MCMC Principles}

MCMC sampling is well-suited for solving high-dimensional and complex probabilistic inference challenges in receiver processing.
The core idea is to simulate a Markov chain whose stationary distribution matches the target distribution (e.g., the posterior distribution of $\mathbf{H}$, $\mathbf{x}$, or $\mathbf{b}$). Each state in this Markov chain represents a sample from the target distribution, and transitions between states are governed by a transition kernel specifically designed to ensure that the chain converges to the target distribution over time.

The most widely used MCMC method is the Metropolis-Hastings (MH) algorithm \cite{brooks2011handbook}. This algorithm works by generating a candidate state from {a simple proposal distribution that encompasses the target distribution \cite{brooks2011handbook}} and then deciding whether to accept or reject this candidate based on a probabilistic rule that ensures the correct stationary distribution is maintained. A special instance of the MH algorithm is Gibbs sampling, where each variable in the state vector is updated sequentially by drawing from its conditional distribution given the other variables. Gibbs sampling is notable for always accepting proposals, making it more efficient in certain scenarios.

\CheckRmv{
  \begin{figure}[t]
    \setlength{\abovecaptionskip}{-0.2cm}
    \centering
    \includegraphics[width=3.3in]{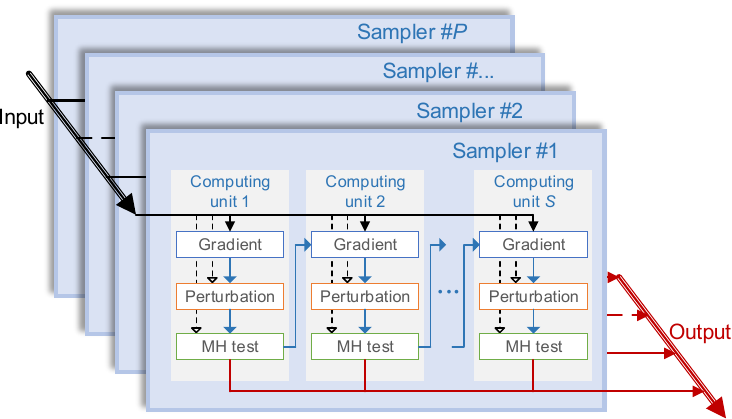}
    \caption{Computing architecture using gradient-based MCMC.} 
    \label{fig:mcmc_computing}
  \end{figure}
}

\vspace{-0.4cm}
\subsection{Gradient-Based MCMC for Receiver Processing}

Recent advances in MCMC algorithms include the use of gradients from the target distribution to guide the sampling process, leading to improved efficiency, especially in non-convex settings where traditional methods struggle \cite{maSamplingCanBe2019}. This gradient-based MCMC approach, exemplified by Langevin Monte Carlo algorithms \cite{cheng2018underdamped}, incorporates gradient information to navigate the landscape of the target distribution more effectively, thereby accelerating convergence.
{In this approach, the target distribution is directly derived from the observation model, requiring only its unnormalized density, thereby eliminating the computational burden of computing the normalizing constant. Furthermore, gradients can be efficiently computed by differentiating the log density, significantly enhancing computational efficiency.} 

In a gradient-based MCMC framework, each proposal generation is determined by a combination of gradient information (which directs the sampling toward high-probability regions) and random perturbations (which prevent the sampling from getting stuck in local minima).
Subsequently, the MH acceptance test is invoked to correct the proposal and steer the sampling toward the intended distribution.
This approach enhances both the speed and accuracy of the sampling process, making it highly suitable for large-scale MIMO systems where computational efficiency is critical.

\figref{fig:mcmc_computing} illustrates the gradient-based MCMC computing architecture, featuring multiple parallel samplers, each corresponding to an independent Markov chain. This parallelism distributes the computational load, improving efficiency and reducing latency. Each sampler consists of basic computing units that perform the sampling process within the MH framework. The combination of gradient information and random perturbations enables efficient state transitions, expediting convergence to the target distribution. Once a sufficient number of samples are obtained, they are used for final inference in tasks such as channel estimation or symbol detection.

In the following sections, we delve into the application of MCMC, particularly the gradient-based MCMC approach, for specific receiver processing tasks, including channel estimation, symbol detection, and channel decoding. Additionally, we explore the potential of MCMC to construct a unified AI receiver architecture---one that optimizes the entire receiver processing pipeline, from estimation to decoding, within a single probabilistic framework.

\vspace{-0.3cm}
\section{MCMC-Based MIMO Receiver Processing}  \label{sec:receiver}

\subsection{Channel Estimation} \label{sec:ce}
Accurate channel state information (CSI) is crucial for leveraging the potential of multiple antennas in MIMO systems, as it directly reflects the signal propagation properties in the communication link. The challenge of channel estimation lies in recovering the CSI given the received signal $\mathbf{y}$ and the known training symbols $\bar{\mathbf{x}}$ (pilots). This task aims to obtain the minimum mean square error (MMSE) estimate, defined as ${\hat{\mathbf{H}} = \mathbb{E}_{\mathbf{H}} [p(\mathbf{H} | \mathbf{y}, \bar{\mathbf{x}})]= \mathbb{E}_{\mathbf{H}}\left[{p(\mathbf{y}|\mathbf{H},\bar{\mathbf{x}}) p (\mathbf{H})}/{p(\mathbf{y})}\right]}$,
where $p(\mathbf{y}|\mathbf{H},\bar{\mathbf{x}})$ represents the likelihood and $p (\mathbf{H})$ the prior distribution that reflects channel statistics.

Recent studies have shown that high-dimensional channels often exhibit inherent structure, typically in the form of sparsity and clustered patterns. These characteristics can be effectively utilized in the estimation process by employing sparse Bayesian learning models. In this approach, the channel coefficients are treated as random variables, with priors specified to encourage sparsity and clustering. Additionally, hyperparameters governing the prior distribution can also be treated as random variables and updated autonomously, forming a complete probabilistic graphical model for channel estimation. The complex posterior inference inherent in this model lends itself to tailored MCMC algorithms.

This concept has been applied to doubly selective multipath channels, where the clustered sparsity of channel taps was considered. A spike-and-slab prior, also known as the Bernoulli-Gaussian process, was used in the Bayesian framework for channel estimation \cite{jingJointChannelEstimation2017}. This model effectively promotes clustered sparsity in the estimated channel coefficients. Due to the intractability of a closed-form posterior estimator, Gibbs sampling was employed for the inference and the acquisition of the channel coefficients. 
{This approach decouples the joint posterior distribution $p(\mathbf{H} | \mathbf{y}, \bar{\mathbf{x}})$ into simpler conditional distributions, allowing iterative updates for each variable.}
A key advantage of this strategy is its ability to update hyperparameters without prior knowledge of the cluster or sparsity levels, making it an unsupervised and practical solution.

Building on this philosophy, we applied a Gibbs sampling scheme to the downlink channel estimation problem in a massive MIMO system with 64 base station (BS) antennas and 8 user terminals. 
{The channels were simulated using the 3GPP spatial channel model, normalized to ensure unit energy per element, and assumed to be quasi-static. Within each transmission frame, 36 time slots were allocated as pilot symbols for channel estimation.} 
A Gaussian mixture prior was adopted to model the channel sparsity due to the limited scattering at the BS. 
{The Gibbs sampling method was executed for 300 iterations before collecting converged samples.
We employed 10 parallel Gibbs samplers, with the final channel estimate obtained by averaging 320 converged samples evenly drawn from these samplers.}

\figref{fig:nmse} presents the normalized mean squared error (NMSE) performance of the channel estimation process. The Gibbs sampling method exhibits significant performance improvements over conventional compressed sensing (CS)-based methods, such as orthogonal matching pursuit (OMP), fast iterative shrinkage-thresholding algorithm (FISTA), and approximate message passing (AMP). 
{Moreover, Gibbs sampling achieves computational complexity comparable to FISTA, while being lower than AMP.}  The performance gain arises from MCMC's ability to perform precise posterior inference. Unlike CS-based methods, which require prior knowledge of the number of significant channel coefficients, or Oracle linear MMSE (LMMSE), which demands exact index identification, the proposed MCMC approach learns these properties adaptively, showcasing a practical advantage in real-world deployments.

\CheckRmv{
  \begin{figure}[t]
    \setlength{\abovecaptionskip}{-0.2cm}
    \centering
    \includegraphics[width=3.0in]{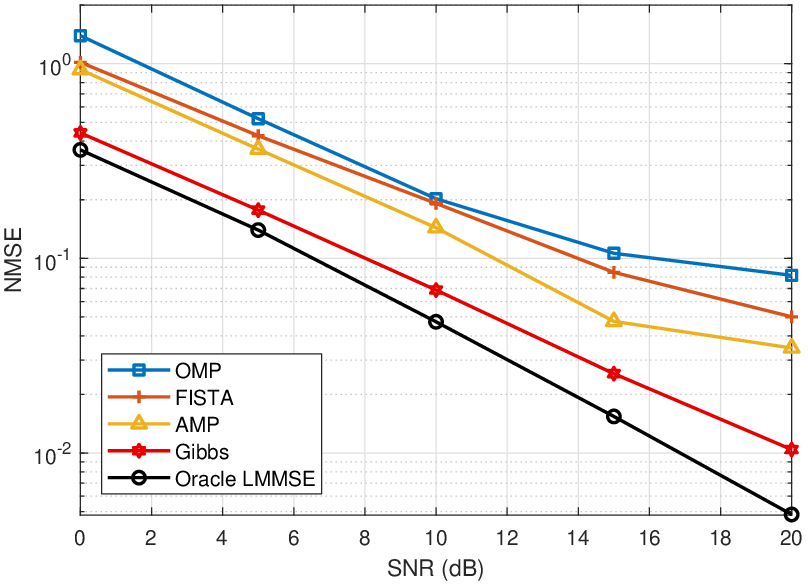}
     \caption{Downlink channel estimation NMSE in a massive MIMO system with 64 BS antennas and 8 user terminals, each equipped with two antennas.}
    \label{fig:nmse}
  \end{figure}
}

\vspace{-0.3cm}
\subsection{Symbol Detection} 
Symbol detection aims to recover the transmitted symbols using the received signal and available CSI. This task represents an integer optimization problem, which is non-deterministic polynomial-time hard due to the discrete nature of the transmitted symbols, typically drawn from a predefined constellation, e.g., quadrature amplitude modulation (QAM). MCMC algorithms offer an efficient solution to this complex problem by reducing the computational burden from exponential levels, required for exhaustive enumeration over the discrete space, to polynomial levels.

The MCMC-based MIMO detector using Gibbs sampling was first developed in \cite{farhang-boroujenyMarkovChainMonte2006}, where samples were generated from the posterior distribution of the transmitted symbols. The Bayesian estimates were then computed from these samples. Several enhancements to this Gibbs-based approach have since been proposed, such as the incorporation of dynamic temperature scaling to increase flexibility and overcome stalling issues at high signal-to-noise ratios (SNRs). This allows the MCMC method to sample from tempered posterior distributions, improving robustness across various SNRs.

A notable advancement in MCMC-based detection comes from the integration of machine learning techniques, such as gradient descent. One example is the use of gradient-based MCMC for MIMO detection, as initially explored in \cite{gowdaMetropolisHastingsRandomWalk2021}. In this approach, MCMC searches the state space by performing random walks guided by the gradient descent direction of the continuous least-squares surface of the objective function. This leads to faster convergence toward high-probability regions. Newton's method and Nesterov's accelerated gradient (NAG) are used in \cite{gowdaMetropolisHastingsRandomWalk2021} and \cite{zhou2023gradient}, respectively, offering a balance between computational complexity and detection performance. 

Along this line of research, a theoretically grounded approach was proposed in \cite{zhou2024near}, introducing a rigorous gradient-based sampling method for discrete spaces. 
Unlike heuristic strategies that fail to implement exact MH correction, this work presents a solution by constructing proposals based on the gradients of the continuous counterpart of the target discrete distribution. The proposal function, restricted to the discrete lattice, is derived using a coordinate-wise factorization of the Gaussian proposal function from the Langevin algorithm. This factorization allows the computational cost to scale quadratically with the number of antennas. With this discrete proposal, exact MH acceptance tests ensure the Markov chain's reversibility and convergence. The final samples, which asymptotically approach the target distribution, are used to compute soft Bayesian estimates of the transmitted symbols. This approach achieves near-optimal detection accuracy while significantly reducing computational overhead.

\CheckRmv{
  \begin{figure}[t]
    \setlength{\abovecaptionskip}{-0.2cm}
    \centering
    \includegraphics[width=3.0in]{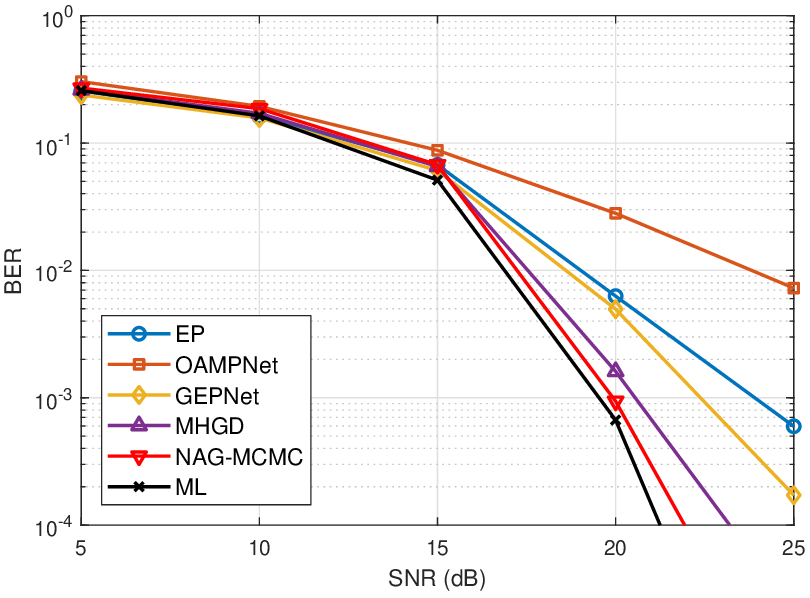}
    \caption{BER performance in an \Times{8}{8} MIMO system with 16-QAM modulation under Rayleigh fading channels.}
    \label{fig:ber_uncoded}
  \end{figure}
}

\CheckRmv{
  \begin{figure}[t]
    \setlength{\abovecaptionskip}{-0.2cm}
    \centering
    \includegraphics[width=3.3in]{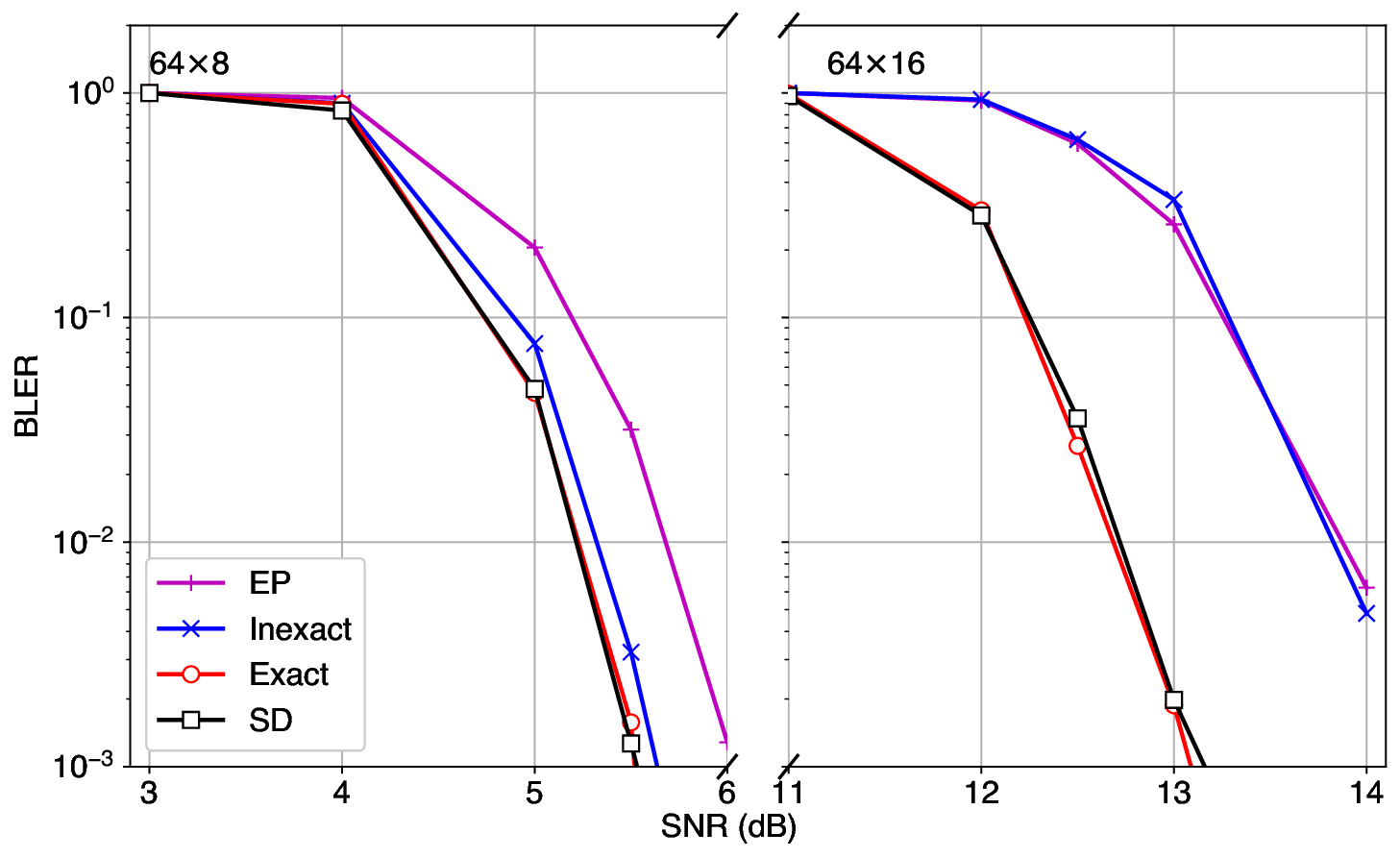}
    \caption{Block error rate (BLER) performance under the 3GPP 3D channel model with 64 BS antennas and 8 or 16 single-antenna users, leading to effective \Times{64}{8} or \Times{64}{16} MIMO setups with 16-QAM modulation.}
    \label{fig:bler}
  \end{figure}
}

\figref{fig:ber_uncoded} compares the bit error rate (BER) performance of MCMC-based detectors against conventional and state-of-the-art NN-based detectors in an \Times{8}{8} MIMO system employing 16-QAM modulation under Rayleigh fading channels. The NAG-aided MCMC (NAG-MCMC) detector from \cite{zhou2023gradient} shows significant improvement over baseline methods such as expectation propagation (EP), deep-unfolding orthogonal AMP networks (OAMPNet), and graph NN-enhanced EP networks (GEPNet). Moreover, its complexity is comparable to that of Newton's method-aided MCMC (MHGD) \cite{gowdaMetropolisHastingsRandomWalk2021}, while being lower than both GEPNet and the optimal ML detector. The results from \cite{zhou2023gradient} also indicate that NAG-MCMC scales efficiently for systems with over 128 antennas.

\figref{fig:bler} presents the performance comparison between inexact and exact gradient-based MCMC methods. An uplink massive MIMO system with 64 BS antennas and either 8 or 16 single-antenna users was simulated using the 3GPP 3D channel model \cite{3gpp36873r12-2017}. 
Both methods used 256 converged samples to compute soft detection. The exact method from \cite{zhou2024near} significantly outperforms the inexact method from \cite{gowdaMetropolisHastingsRandomWalk2021} and delivers near-optimal performance, comparable to sphere decoding (SD)-based detectors, which are computationally expensive. The exact gradient-based MCMC method proves to be computationally efficient, particularly for large-scale systems, with only a minimal number of crucial samples required to achieve competitive performance.

\vspace{-0.4cm}
\subsection{Channel Decoding}
Channel decoding is responsible for recovering the raw message bits using information about the estimated transmitted symbols and the applied coding scheme. This process is crucial in modern communication systems, as it significantly enhances the reliability and error rate performance of data transmission.

For forward error correction, channel decoders typically use either the generator or parity check matrices that define the coding principle as input. MCMC-based methods can be applied to decode the transmitted bits by sampling from the posterior distribution of the message bits. As the Markov chain converges, the final message estimate is derived either through statistical averaging or bit counting, producing the estimated bits or their log-likelihood ratios (LLRs). This MCMC-based approach offers a distinct advantage over traditional decoders, as it is guaranteed to approach information-theoretic optimality with polynomial complexity \cite{huang2023parallel}, making it a promising area for further research.

The use of MCMC in decoding low-density parity-check (LDPC) codes was first explored by Radford Neal in 2001 \cite{nealMonteCarloDecoding2001}, where a Gibbs sampling-based decoder was developed. However, this approach faced significant challenges due to its slow convergence speed. With the advancement of MCMC techniques and numerical computing, interest in MCMC-based decoding has seen a resurgence. Much of the recent research focuses on applying MCMC methods to short codes, which addresses some of the drawbacks of the deterministic belief propagation (BP)-based decoding, such as its sensitivity to loops in the factor graph.

A common approach is to use the MH algorithm, where only one bit is flipped per iteration for proposal generation. The error syndrome of the bit sequence serves as the measure for computing the acceptance probability of the proposed sample. As the algorithm proceeds, samples distributed according to the posterior distribution of the bit sequence are generated, resulting in a theoretical error rate that is less than twice the optimal MAP decoding error rate.

{Another widely used method is Gibbs sampling, where at each iteration, the full conditional probability of flipping a particular bit is calculated and used for updating the bit.
This method converges to sampling from the joint posterior distribution of the entire code sequence.}  
Further enhancements, such as simulated annealing (which dynamically adjusts the Markov chain's temperature) and bit update order rearrangement, have been introduced, leading to significant performance improvements over traditional BP-based LDPC decoding.

Despite the advancements, one limitation of existing MCMC-based decoding schemes is their confinement to the short code domain, primarily due to the inefficiency of basic Gibbs and MH methods. Nonetheless, similar to the developments seen in MCMC-based symbol detection, gradient-informed proposals can be leveraged to guide the sampling process in decoding tasks. This has led to a growing body of research aimed at integrating gradient-based proposals into discrete sampling schemes like Gibbs sampling, significantly improving their scalability for larger codes. As a result, this approach has the potential to expand the application of MCMC-based decoding beyond short codes, addressing the limitations that also hinder many machine learning or NN-based decoding methods.

\section{MCMC-Based Unified Receiver Architecture} \label{sec:unified}
\CheckRmv{
  \begin{figure*}[t]
    \setlength{\abovecaptionskip}{-0.1cm}
    \centering
    \includegraphics[width=6.1in]{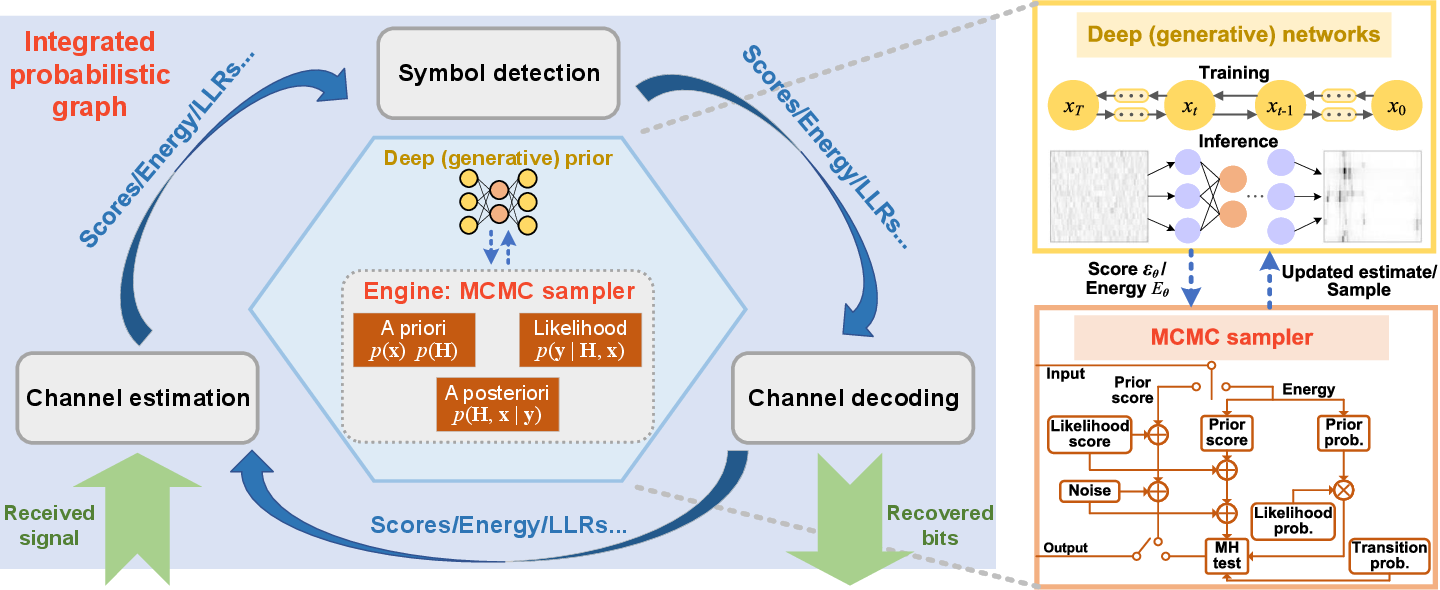}
    \caption{Diagram of the MCMC-based unified receiver architecture. The left part illustrates the overall framework, while the right part details the integration of MCMC with data-driven NNs.}
    \label{fig:joint}
  \end{figure*}
}

In addition to functioning as a solver for individual processing modules, MCMC also enables a universal architecture that unifies the receiver's processing tasks, as shown in \figref{fig:joint}. This architecture accepts the received signals as input and outputs the recovered data (message bits). All receiver processing tasks are formulated within a Bayesian statistical learning framework, with MCMC sampling serving as the core engine. The key concepts and advantages of this architecture are described in the following subsections.

\subsection{Integrated Architecture for Receiver Processing and Optimization}
The proposed architecture employs MCMC as a universal Bayesian computing engine to standardize core receiver processing tasks---channel estimation, symbol detection, and channel decoding---all formulated as posterior inference problems. Fundamental gradient-based MCMC blocks, such as those illustrated in \figref{fig:mcmc_computing}, can provide a generic solution for these tasks.
Additionally, \figref{fig:joint} shows their integration into a single probabilistic graph for joint performance optimization. Unlike black-box approaches, the framework retains modular processing blocks that iteratively refine intermediate estimates through MCMC-guided information exchange, enabling flexible performance-complexity tradeoffs.

\subsection{Hybrid Data- and Model-Driven Learning}
The unified receiver architecture also facilitates the seamless integration of data-driven and model-driven approaches, forming a hybrid learning system (illustrated in the right part of \figref{fig:joint}). A prominent example of this integration is the use of deep generative networks to model the channel prior $p(\mathbf{H})$, which replaces conventional handcrafted priors (as discussed in \secref{sec:ce}). These data-driven models effectively capture complex propagation characteristics without requiring explicit knowledge of the channel distribution; this is particularly valuable in high-dimensional, nonlinear scenarios where traditional channel models fail.

Diffusion models (DMs) and score-based generative models \cite{arvinteMIMOChannelEstimation2022} exemplify this generative approach. Once trained on large-scale channel datasets, these models can employ either a score-based or an energy-based parameterization \cite{du2023reduce} to characterize the underlying channel distribution. In the score-based method, a NN is used to learn the gradient of the logarithm of the channel distribution, referred to as ``scores.'' These scores, combined with the likelihood scores derived from observations, are utilized in annealed Langevin dynamics---a gradient-based MCMC variant---for efficient posterior sampling from high-density regions.

Alternatively, the energy-based parameterization provides complementary advantages by explicitly evaluating the unnormalized densities of the channel prior, which allows for exact MCMC sampling and posterior inference. This approach facilitates MH corrections within the MCMC framework by incorporating energy changes in the proposed samples, likelihood densities, and transition probabilities. In addition, scores can be derived from the gradient of the log density, enabling Langevin-based exploration similar to the score-based approach. The resulting Metropolis-adjusted sampling, supported by pre-trained energy-based DMs, further enhances posterior inference accuracy \cite{du2023reduce} and improves channel acquisition performance.

Both parameterizations empower deep generative models to serve as accurate, data-driven proxies for the channel, thereby enabling end-to-end receiver optimization. Furthermore, dynamic updates of the scores or energy representations based on real-time channel realizations enhance inter-module interactions beyond conventional LLR information exchange. This synergy between data-driven learning and model-driven MCMC inference establishes a universal AI-driven receiver that adapts to complex, dynamic channel conditions while generalizing across diverse communication scenarios.

\CheckRmv{
  \begin{figure}[t]
    \setlength{\abovecaptionskip}{-0.2cm}
    \centering
    \includegraphics[width=3.0in]{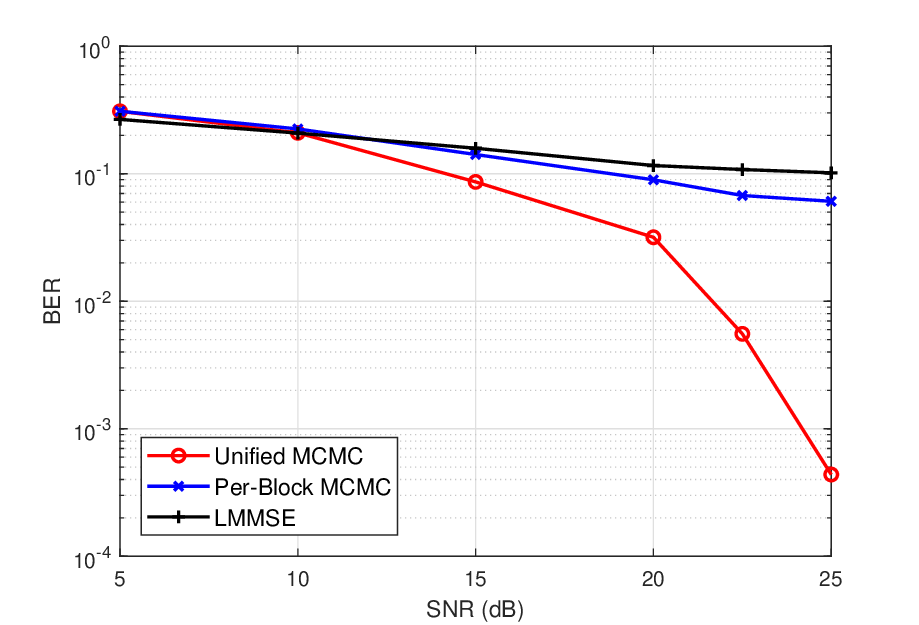} 
    \caption{Performance of the MCMC-based unified receiver under the 3GPP 3D channel model with 64 BS antennas and 32 single-antenna users, yielding effective \Times{64}{32} MIMO setups with QPSK modulation.} 
    \label{fig:ber_unified}
  \end{figure}
}

\subsection{Performance Evaluation}
The efficacy of the unified receiver design is demonstrated in \figref{fig:ber_unified} through the joint optimization of channel estimation and symbol detection. We consider an uplink MIMO system under the 3GPP 3D channel model \cite{3gpp36873r12-2017}, featuring a BS with 64 antennas serving 32 single-antenna users employing quadrature phase shift keying (QPSK) modulation. Detailed simulation setups for channel generation are provided in \cite[Section IV-E]{zhou2024near}. In a block-fading scenario with an 80-slot coherence time (comprising 30 pilot slots and 50 data slots), the unified MCMC approach leverages a pre-trained energy-based DM to represent channel priors while employing the Metropolis-adjusted Langevin algorithm to sample from the joint posterior distribution of channels and symbols. This is achieved through an iterative process that alternates between the channel estimation and symbol detection modules, with each module updating its variables via gradient-based MCMC and exchanging information to refine the estimates. The results shown in \figref{fig:ber_unified} indicate that our approach significantly outperforms the traditional LMMSE channel estimation combined with symbol detection, as well as the per-block MCMC method described in \secref{sec:receiver} that processes channel estimation and symbol detection separately.


\section{Future Research Directions}  
Although MCMC-based MIMO receiver processing holds great promise, it is still in the early stages of development. Several challenges must be addressed to achieve an efficient, unified, and autonomous AI-driven receiver framework that deeply integrates MCMC and communications. This section outlines some promising avenues for future research.

\subsection{Customized MCMC Algorithms for Communications} 

MCMC methods, originating from the fields of physics and statistics, offer accurate inference for complex models but have not yet been widely explored within the communications domain. Customizing MCMC algorithms for communication signal processing and establishing the corresponding theoretical guarantees require immediate attention. As communication networks grow in complexity, leading to heterogeneous data and signals, receiver parameters often exhibit diverse and mixed properties, combining discrete and continuous variables.
{Advancing discrete sampling methods and MCMC algorithms tailored to hybrid distributions, incorporating both discrete and continuous components, along with refining theoretical foundations, represents a promising direction for future research.} 

\subsection{MCMC Algorithms and Architectures for Distributed MIMO Processing}  
Centralized MIMO processing encounters significant roadblocks as antenna arrays become larger, with increased processing complexity and interconnection bandwidth demands. Distributed signal processing is becoming imperative for ultra-massive MIMO systems, but current MCMC schemes are primarily designed for centralized architectures. Given MCMC's native support for parallel computing, it has the potential to excel in distributed processing by efficiently integrating and allocating computational resources while minimizing data exchange between distributed units. The development of decentralized MCMC algorithms, distributed computing architectures, and supporting theoretical frameworks will be key to unlocking MCMC's potential in distributed MIMO systems.

\subsection{General-Purpose Bayesian Computing Engine and Hardware Design}
Extending the MCMC-based unified receiver architecture discussed in \secref{sec:unified}, there is an opportunity to create general-purpose Bayesian computing engines. These engines could handle not only communication signals but also heterogeneous data from broader services and infrastructure within communication networks. This evolution requires efficient MCMC-based processing schemes, standardized computing procedures, and robust theoretical foundations. Furthermore, the design of hardware accelerators tailored to communication tasks is essential for improving computational and energy efficiency. Hardware-software co-design, which includes coordinating between statistical/AI computing toolchains and specialized hardware, will play a critical role in achieving these goals.

\subsection{Deep Integration of MCMC and Communications}
A key challenge for future research is the effective integration of customized MCMC algorithms, generalized architectures, and specialized hardware accelerators into a cohesive system. Achieving deep integration between MCMC and communications requires a holistic approach that seamlessly combines these components. A promising direction involves leveraging both expert knowledge and AI capabilities to embed MCMC as a service within communication networks. This approach aligns with the 6G vision of deeply integrated AI and communications, enabling joint optimization of core communication tasks and diverse external applications such as sensing and AI-driven functionalities. By capitalizing on commonalities between these tasks, this integration can push performance boundaries and enable highly autonomous communication systems.

\section{Conclusion}  
In this paper, we explored the use of MCMC methods in MIMO receiver processing and proposed a unified, AI-driven receiver architecture that leverages MCMC as its core computing engine. This integration presents a promising avenue for realizing a deep synergy between AI and communication systems. The proposed approach not only addresses the immediate challenges of MIMO receiver processing but also aligns with the broader objective of embedding AI as a fundamental service within future wireless networks. Through this framework, MCMC and similar AI methodologies are envisioned to evolve from being mere problem solvers into central enablers of an adaptive, intelligent, and cohesive communication infrastructure that can meet a diverse range of demands.



\ifCLASSOPTIONcaptionsoff
  \newpage
\fi



\begin{thebibliography}{10}
  \providecommand{\url}[1]{#1}
  \csname url@samestyle\endcsname
  \providecommand{\newblock}{\relax}
  \providecommand{\bibinfo}[2]{#2}
  \providecommand{\BIBentrySTDinterwordspacing}{\spaceskip=0pt\relax}
  \providecommand{\BIBentryALTinterwordstretchfactor}{4}
  \providecommand{\BIBentryALTinterwordspacing}{\spaceskip=\fontdimen2\font plus
  \BIBentryALTinterwordstretchfactor\fontdimen3\font minus
    \fontdimen4\font\relax}
  \providecommand{\BIBforeignlanguage}[2]{{%
  \expandafter\ifx\csname l@#1\endcsname\relax
  \typeout{** WARNING: IEEEtran.bst: No hyphenation pattern has been}%
  \typeout{** loaded for the language `#1'. Using the pattern for}%
  \typeout{** the default language instead.}%
  \else
  \language=\csname l@#1\endcsname
  \fi
  #2}}
  \providecommand{\BIBdecl}{\relax}
  \BIBdecl

  \bibitem{wang2024tutorial}
  Z.~Wang \emph{et al}., ``A tutorial on extremely large-scale MIMO for 6G:
  Fundamentals, signal processing, and applications,'' \emph{IEEE
  Commun. Surveys Tuts.}, vol. 26, no. 3, pp. 1560--1605, 3rd Quart., 2024.

  \bibitem{qin2024ai}
  Z. Qin, \emph{et al}., ``AI
  empowered wireless communications: From bits to semantics,'' \emph{Proc.
  IEEE}, vol. 112, no. 7, pp. 621--652, Jul. 2024.

  \bibitem{brooks2011handbook}
  S.~Brooks, A.~Gelman, G.~Jones, and X.-L. Meng, \emph{Handbook of {{Markov Chain Monte Carlo}}.}
    \hskip 1em plus 0.5em minus 0.4em\relax Boca Raton, FL, USA: {Chapman \& Hall}, 2011.

  \bibitem{maSamplingCanBe2019}
  Y.-A. Ma, Y.~Chen, C.~Jin, N.~Flammarion, and M.~I. Jordan, ``Sampling can be
    faster than optimization,'' \emph{{Proc. Nat. Acad. Sci.}}, vol. 116, no.~42, pp. 20\,881--20\,885, Sep. 2019.

  \bibitem{cheng2018underdamped}
  X. Cheng, N. S. Chatterji, P. L. Bartlett, and M. I. Jordan, ``Underdamped
  Langevin MCMC: A non-asymptotic analysis,'' in \emph{Proc. Conf. Learn.
  Theory (COLT)}, 2018, pp. 300--323.

  \bibitem{jingJointChannelEstimation2017}
  L.~Jing, C.~He, J.~Huang, and Z.~Ding, ``Joint channel estimation and detection
    using {{Markov}} chain {{Monte Carlo}} method over sparse underwater acoustic
    channels,'' \emph{IET Commun.}, vol.~11, no.~11, pp. 1789--1796, Sep. 2017.

  \bibitem{farhang-boroujenyMarkovChainMonte2006}
  B.~{Farhang-Boroujeny}, H.~Zhu, and Z.~Shi, ``Markov chain {{Monte Carlo}}
    algorithms for {{CDMA}} and {{MIMO}} communication systems,'' \emph{IEEE
    Trans. Signal Process.}, vol.~54, no.~5, pp.~1896--1909, May 2006.

  \bibitem{gowdaMetropolisHastingsRandomWalk2021}
  N.~M. Gowda, S.~Krishnamurthy, and A.~Belogolovy, ``Metropolis-{{Hastings
    random walk}} along the {{gradient descent direction}} for {{MIMO
    detection}},'' in \emph{Proc. {{IEEE Int. Conf.}}
    {{Commun.}} ({{ICC}})}, Jun. 2021, pp. 1--7.

  \bibitem{zhou2023gradient}
  X.~Zhou, L.~Liang, J.~Zhang, C.-K. Wen, and S.~Jin, ``{{Gradient-based Markov chain Monte Carlo for MIMO detection}},'' {\emph{IEEE Trans. Wireless Commun.}, vol.~23, no.~7, pp. 7566--7581, Jul. 2024}.

  \bibitem{zhou2024near}
  {X.~Zhou, L.~Liang, J.~Zhang, C.-K. Wen, and S.~Jin, ``Near-optimal {MIMO} detection using gradient-based {MCMC} in discrete
    spaces,'' \emph{IEEE Trans. Signal Process.}, vol. 73, pp. 584--600, 2025.}

  \bibitem{3gpp36873r12-2017}
  {3GPP TR 36.873}, ``Study on 3D channel model for LTE ({Release} 12),'' Tech. Rep., Jun. 2017. [Online]. Available:
    \url{https://www.3gpp.org/ftp/Specs/archive/36_series/36.873/}.


  \bibitem{huang2023parallel}
  J.-T. Huang and Y.-H. Kim, ``Parallel Monte Carlo Markov chain decoding of linear codes,'' in \emph{Proc. IEEE Int. Symp. Inf. Theory (ISIT)}, Jun. 2023, pp. 2051--2056.


  \bibitem{nealMonteCarloDecoding2001}
  R.~M. Neal, ``Monte {{Carlo}} decoding of {{LDPC}} codes,'' {Dept. of Computer Science, University of Toronto}, Tech. Rep., 2001.

  \bibitem{arvinteMIMOChannelEstimation2022}
  M.~Arvinte and J.~I. Tamir, ``{{MIMO channel estimation}} using {{score-based
  generative models}},'' \emph{IEEE Trans. Wireless Commun.},
  vol.~22, no.~6, pp. 3698--3713, Jun. 2023.

  \bibitem{du2023reduce}
  {Y. Du \emph{et al.}, ``Reduce, reuse, recycle: Compositional generation with energy-based diffusion models and MCMC,'' in \emph{Proc. Int. Conf. Machine Learning (ICML)}, 2023, pp. 8489--8510.}

\end{thebibliography}


\end{document}